\documentclass[prb,preprint]{revtex4-1} 

\usepackage{amsmath}  
\usepackage{amsfonts} 
\usepackage{graphicx} 

\begin{document}


\title{Magnetic Solitons for Non Heisenberg Anisotropic Hamiltonians in Linear Quadrupole Excitations}

\author{Yousef Yousefi}
\email{Yousof54@yahoo.com} 
\altaffiliation[permanent address: ]{Payame Noor University
 POB 19395-4697,Tehran,Iran.  
  } 
\affiliation{Department of Physics, Payame Noor University, Tehran, IRAN}



\date{\today}

\begin{abstract}
We discuss system with non-isotropic non-Heisenberg Hamiltonian with nearest neighbor exchange within a mean field approximation process. We drive equations describing non-Heisenberg non-isotropic model using coherent states in real parameters and then obtain dispersion equations of spin wave of dipole and quadrupole branches for a small linear excitation from the ground state. In final, soliton solution for quadrupole branches for these linear equations obtained.
\end{abstract}

\maketitle 

\section{Introduction} 

During the past decade study of nonlinear behavior of magnetic crystals has been attracted large attention, specially it accompany with the progress in some other fields such as development of theory of nonlinear differential equation, achieving new laboratory results and also potential applications in other branches of science and technology [1, 2].

Particles with spin  $S\geq 1$  are more interesting among the other nano particles [3, 4].  This is because of existing of complexity in their behavior due to their multipole dynamic spin excitations. In such systems, the number of necessary parameters for complete description of macroscopic properties increases up to 4S, that S stands for magnitude of system spin.

Also it worthwhile, the process of achieving classical spin equations and dynamic multipoles is based on coherent states that are obtained in $SU(2S+1)$  group[5].

We consider unitary anisotropic Hamiltonian as form of:

\begin{eqnarray}
\hat H=-J\sum_i(\hat S_i \hat S_{i+1}+\delta\hat S_i^z \hat S_i^z)
\end{eqnarray}

Which, $ \hat S_i^x , \hat S_i^y ,\hat S_i^z$   are  spin operators in lattice $i$, and $\delta$  is  anisotropy coefficient. This is Hamiltonian of one dimensional ferromagnetic spin chain observed in compositions like $CSNiF_3$ [6].

In this paper the goal is to obtain classical equation for stated Hamiltonian and finding the answer of spin wave for small linear excitations upper than the ground state. Coherent states issued nearest approximation to classical state i.e. pseudo classical, because they minimize uncertainty principles. For this reason, in section 2, coherent states for spin $S=1$ developed that are the same as coherent states in SU(3) group. To obtain classical Hamiltonian, we need average values of spin operator; so in section 3, these values and classical Hamiltonian equation are derived. In section 4, Hamiltonian equation computed in previous section is substituted in classical equations of motion resulted from using Feynman path integral on coherent states, and then we acquire spin wave equations and dispersion equations of dipole and quadupole branches for small linear excitation above the ground state, and finally we calculate soliton answers of linearized equations. 

\section{Coherent states in SU(3) group}

Coherent states are special quantum states that their dynamic is very similar to behavior of their classical system. The kind of coherent state that is used in a problem depends on symmetry of existent operators. With considering existent symmetry in operators of Hamiltonian (1), coherent states in SU(3) group is used for accurate description and considering all multipole excitations. In this group, ground state considered as $(1,0,0)^T$ and its single-site coherent state is written as:

\begin{eqnarray}
|\psi \rangle = D^1(\theta, \phi)e^{-i\gamma \hat S^z} e^{2ig\hat Q^{xy}}|0\rangle 
\end{eqnarray}

In above equation, $D^1 (\theta,\phi)$ is Wigner function for spin $S=1$ and two angles $\theta$ and $\phi$ determine alignment of classical spin vector. Angle $\gamma$ determines direction of quadruple torque around the spin vector. Parameter g specifies change of length of average value of quadruple torque and also of magnitude of spin vector. Lagrangian can be obtained by use of Feynman path integral for declared coherent states as:[8]

\begin{eqnarray}
L=cos2g(cos\theta \phi_t -\gamma_t )-H(\phi, \theta, g, \gamma)
\end{eqnarray}

Where $ x_t=\frac{\partial}{⁄\partial t}$  and H is classical energy of system obtained by averaging Hamiltonian (1) on coherent states (2).

Two other terms appear when acquiring Lagrangian of spin system. The first is Kinetic term that has Berry phase characteristics issued from quantum interference of Instanton paths and has important role in quantum phenomenons such as spin tunneling and the second is boundary term that depends on boundary values of path. Both of term have no role in classical dynamic of spin excitations and so are not considered here.

\section{Classical Hamiltonian and equations in SU(3) group}

Average spin values in SU(3) group written as:[9]

\begin{eqnarray}
S^+ &=& e^{i\phi}cos(2g)sin\theta  \nonumber\\
S^- &=& e^{-i\phi} cos(2g)sin\theta \nonumber\\
S^z &=& cos(2g) cos\theta 
\end{eqnarray}

By averaging Hamiltonian (1) and using (4), the continuous limit of classical Hamiltonian obtained as:[4]

\begin{eqnarray}
H_{cl}&=& -J\int \frac{dx}{a_0}(cos^2( 2g)+\frac{\delta}{2}(cos^2\theta+sin(2g)cos(2\gamma) sin^2\theta) \nonumber\\
&  &-\frac{a_0^2}{2}((\theta_x^2+\phi_x^2 sin^2\theta)cos^2( 2g)+4g_x^2 sin^2( 2g)))
\end{eqnarray}

To obtain classical equation of motion, the above classical Hamiltonian is substituted in motion equations resulted from Lagrangian equation:

\begin{eqnarray}
\frac{1}{\omega_0}\phi_t &=& \delta cos\theta(sec(2g)-cos(2\gamma) tan(2g))+a_0^2cos(2g)(\theta_{xx}csc\theta+\phi_x^2cos\theta) \nonumber\\
\frac{1}{\omega_0}\theta_t &=& \frac{\delta}{2}sin(2\theta) sin(2\gamma) tan(2g)-a_0^2 \phi_{xx}cos(2g)sin\theta \nonumber\\
\frac{1}{\omega_0}g_t &=&- \frac{\delta}{2}sin(2\gamma) sin^2\theta  \nonumber\\
\frac{1}{\omega_0}\gamma_t &=&(4cos(2g)-\delta(cos(2\gamma)(cot(4g)-cos(2\theta) csc(4g))+cos^2\theta sec(2g) )) \nonumber\\
&  &+(cos(2g)(8g_x^2-2\theta_x^2+\frac{1}{2}\phi_x^2(-3+cos(2\theta))-\theta_{xx}cot\theta)+4g_{xx}sin(2g))a_0^2 \nonumber\\
&   &
\end{eqnarray}

These equations completely describe nonlinear dynamics of Hamiltonian of problem up to quadrupole excitation. Solutions of these equations are magnetic solitons. These equations result Landau-Lifshitz equation if quadrupole excitations ignored $(g=0)$. So these equations are more general in comparison with landau-Lifshitz and have more degree of freedom. It’s noteworthy that solution of these equations has different range of solitons. 

  For small linear excitation from ground stste, classical equations of motion change to:

\begin{eqnarray}
\frac{1}{\omega_0}\phi_t &=& \delta (secg_0+tang_0)\theta+a_0^2cosg_0\theta_{xx} \nonumber\\
\frac{1}{\omega_0}\theta_t &=& -a_0^2 \phi_{xx}cosg_0 \nonumber\\
\frac{1}{\omega_0}g_t &=&- \frac{\delta}{2}\gamma  \nonumber\\
\frac{1}{\omega_0}\gamma_t &=&-2(2sing_0+\frac{\delta}{cosg_0})g+4a_0^2 g_{xx}sing_0 \nonumber\\
\end{eqnarray}

To obtain dispersion equations, functions $\theta, \phi ,\gamma$ and g are considered as plane waves and their substitution in linearized equations result in dispersion equation for spin wave near the ground state:

\begin{eqnarray}
\omega_1^2 &=& \omega_0^2 k^2 a_0^2 (\delta(1+sing_0)+k^2 a_0^2 cos^2g_0) \nonumber\\
\omega_2^2 &=& \omega_0^2 [2sing_0 k^2 a^2_0+\delta (\frac{4\delta}{sin^2g_0}-2sing_0)]
\end{eqnarray}

From the above equation, it is obvious that both dipole and quadruple branches of unitary Hamiltonian are dispersive in presence of linear excitations.

To compute soliton answers of equations (7), we define variable $\eta$ such as $\eta=x-vt$. In this case above equations convert to below nonlinear equations.

\begin{eqnarray}
\frac{v^2}{\omega_0^2}+\delta a_0^2 (1+sing_0)\theta_{\eta}+( a_0^2 cosg_0)^2\theta_{\eta \eta \eta}=0 \nonumber\\
\frac{-2}{\delta\omega_0}g_{tt}+2(2sing_0+\frac{\delta}{2cosg_0})g-4a_0^2g_{xx}sing_0=0
\end{eqnarray}

The first equation is third order differential equation. So change of dipole moment in Hamiltonian (1) is not of the form of soliton. Solution of this equation has the following forms:

\begin{eqnarray}
\theta=C sin[(x-vt)(\frac{(a_0^2cosg_0)^2}{\frac{v^2}{\omega_0^2}+\delta a_0^2 (1+sing_0)})^{1/2}]
\end{eqnarray}

The second equation is nonlinear Klein-Gordon equation and shows change of average value of quadruple excitation that its solution is of the form of Hylomorphic solitons. These solitons are like Q-ball solitons. The reason of this name is because of they cause matter have appropriate form. Also these solitons are of the kind of non topologic ones because their boundary values in ground and infinity are the same from the topological point of view. If rewrite nonlinear Klein-Gordon equation (9) as:

\begin{eqnarray}
g_{tt}=\alpha g_{xx}+\beta g
\end{eqnarray}

Where 

\begin{eqnarray}
\alpha &=&-\delta \omega_0a_0^2sing_0 \nonumber\\
\beta&=&\frac{\delta \omega_0(4sin2g_0+\delta)}{8cosg_0}
\end{eqnarray}

 Numerical solution of (11) is plotted in figure (1). In this computation we consider $\alpha=10^5$  and  $\beta=10^{10}$ . 

\begin{figure}[htb]
\centerline{\includegraphics[width=0.8\textwidth]{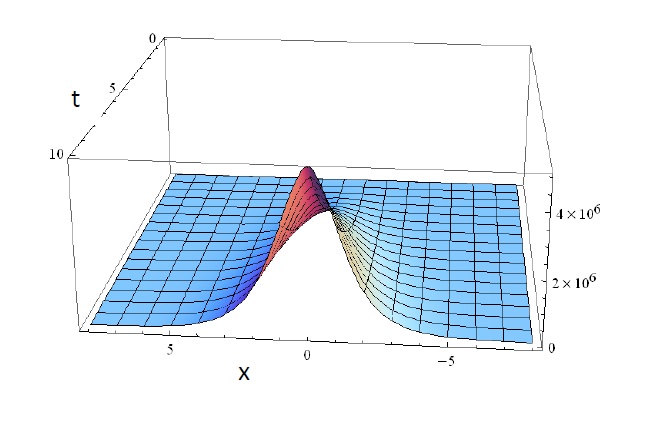}}
\caption{Numerical solution (quadruple excitation) of relation (12) is Helomorphic soliton. }
\end{figure}

Analytical solution of above nonlinear Klein-Gordon equation is the following form:

\begin{eqnarray}
g(x,t)=C sinh[(x-vt)(\frac{-\omega_0(4sin2g_0+\delta)}{cosg_0(v^2+\delta \omega_0a_0^2sing_0)})^{1/2}]
\end{eqnarray}

Where C is constant.

\section{Conclusion}

In this paper, we study semi-classic theory for spin systems with spin $ S=1$ that contain anisotropic exchange terms. it is shown that for anisotropic ferromagnet, value of average quadruple torque is not constant  ($g_t\neq0$)    and its dynamic contains rotational term around classical spin vector ($\gamma_t\neq0$)  and another dynamics that relates to change of length of quadruple torque. There are no such excitations in regular magnets and their dynamics is achieved by use of average value of Heisenberg spin Hamiltonian. Also it is shown that soliton solutions are of the kind of non topologic Hilomorphic solitons for quadruple excitations.

\end{document}